# Building a Simplistic Automatic Extruder: Instrument Development Opportunities for the Laboratory

Stefanie Klisch,[1] Dylan Gilbert,[2] Emma Breaux,[1] Aliyah Dalier,[2] Sudipta Gupta,[1] Bruno Jakobi,[1] Gerald J. Schneider[1,3]

1 Department of Chemistry, Louisiana State University, Baton Rouge, LA 70803

2 Department of Chemistry and Physics Southeastern Louisiana University, Hammond, LA 70402

3 Department of Physics & Astronomy, Louisiana State University, Baton Rouge, LA 70803

## ABSTRACT

A well-rounded introduction to work in a STEM laboratory is vital to scientific education. Besides the ability to use available instrumentation for sample characterization, students should also be imparted knowledge in the steps of instrument development and construction. These concepts can be taught using the example of lipid vesicle preparation via extrusion. Vesicle extrusion is a common technique that involves syringes pushing solutions through membrane filters and is used in fundamental studies on vesicles. Such research is important to better understand of biological phenomena and drug development. Well prepared samples are key to successful research. While the manual approach is very useful to acquire experience, automatic extrusion is more convenient, and automation often results in better reproducibility. These advantages can be combined in a simplistic automatic extruder, that does not require advanced technical skills to be assembled. It can therefore be used by various groups, ranging undergraduate to graduate students using equipment typically available. Using this approach, students can acquire different skillsets including coding, testing, and advanced use of building materials based on their properties. Finally, the quality of the automatic extruder is verified.

## GRAPHICAL ABSTRACT

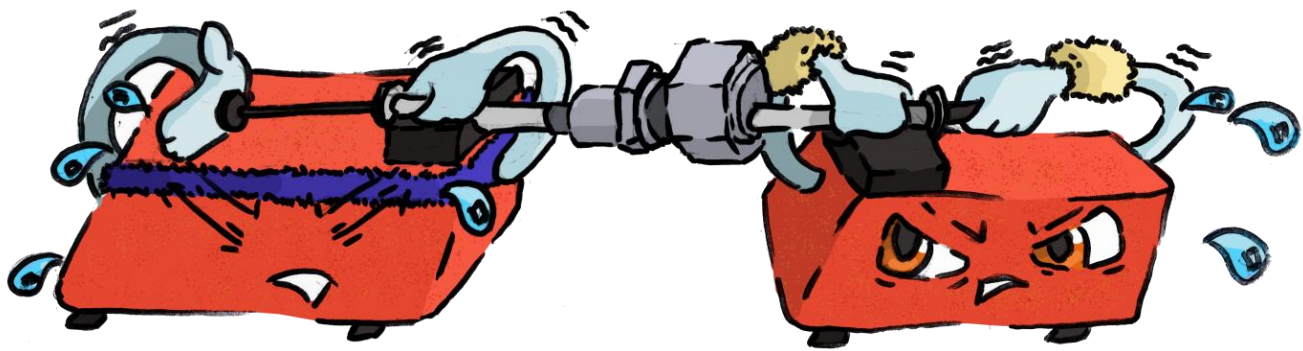

## INTRODUCTION

First described in 1964,[1] liposomes have become an important asset to modern research. The recent boom in gene-therapies now takes advantages of the inner aqueous core of vesicle systems like liposomes to transport medical cargo.[2] Further applications include cosmetics[3, 4] medical imaging[5, 6] and vaccines.[7, 8]

Basic research projects taking advantage of these soft-shelled self-assemblies often begin with the preparation of liposomes. Several techniques are available, including electroporation,[9] sonication,[10] the temperature switch method,[9] the use of a caterpillar mixer[11] and the double emulsion technique.[12] Due to the simplicity of use and the ability to control vesicle radius, extrusion has become a common method.[10, 13] Extrusion relies purely on the physical passage of the vesicles through the pores of the membrane, without the need of additives that may change liposome properties.[14-16]

Simple hand extruders consist of two glass syringes, membranes, and support structures. Commercial tools like the Avanti Mini-extruder,[17] the Liposofast Liposome Factory,[18] the Twist by Helix Biotech[19] or the Genizer Hand Extruder[20] are pre-configured sets that can be used to manually extrude amphiphilic molecules to obtain vesicles. 15 or more extrusion passes are common to create specific size and size uniform liposomes.[21] Hence, the process of preparing multiple samples can be long, monotonous, and exhausting, posing a problem for accessibility.

Automatic extrusion is an attractive alternative, and several different possibilities are commercially available.[13, 14, 16, 22-25] However, automatic extruders are often expensive and complex. They may require custom machinery[13] or necessitate students to work with pressurized gas.[15, 23, 24, 26-29] The following is a procedure to build an automatic extruder using affordable parts that are often available in a typical chemistry laboratory. The assembly and testing of a minimalistic system are described, and a software based on LabVIEW introduced. In this way, student researchers can build their own device.

The current lack of free software matching this project presents a barrier. Hence, a software, based on LabVIEW, was developed, and licensed under the MIT License. It is freely available and can be modified, copied, and updated versions can be re-uploaded.[30] Due to its graphical user interface, LabVIEW provides an excellent approach for visual software development. [31-35] Basic knowledge of LabVIEW is provided in entry level textbooks.[36-39]

The automatic extruder is then compared to manual extrusion. For that purpose 1,2-Dioleoyl-sn-glycero-3-phosphocholine (DOPC), a common,[40] safe,[41] and well-studied amphiphile that is known to form vesicles when prepared by extrusion was used. DOPC consists of a negatively charged phosphatidyl group, a positively charged choline and two oleoyl tails.[40] Because phosphatidylcholines remain zwitterionic over a large pH range,[42] their behavior is unlikely to be affected by small changes in pH. No buffer is necessary and the water available in teaching laboratories can be used (**Figure SI 1**). The automatic extruder is shown to work reliably. When employed as described here, extrusion of self-assembled vesicular systems can be a part of a macromolecular chemistry lab demonstrating a process that is also used in industry.[43]

Models describing the process of extrusion through a membrane are available in literature. These models generally assume laminar flow (Q) within the pore.[44, 45] Clerc et al. proposed a model where a velocity gradient within the pore causes shear of the individual lipid lamellae.[44] The fluid near the center of the pore moves faster than near the edge of the pore. As large vesicles entering the pore are bent to a curvature smaller than what their bilayer can tolerate, they break and re-seal into smaller vesicles. Since the movement of the solution within the pore is laminar, Darcy's Law has been widely used to describe the extrusion process.[23, 44, 45] According to this law, the flow rate shows a linear dependence on the pressure difference between the entrance and exit of the pore.[23, 46]

$$\eta Q = K \left( \frac{\Delta P}{\Delta L} \right) \tag{1}$$

Here, $Q$ is the rate of laminar flow as it passes through a cylindrical pore. Over a distance $\Delta L$, there is an applied pressure difference of $\Delta P$. Darcy's Law describes the behavior of a homogenous fluid. Applied to solutions containing vesicles, it holds true for extrusions at high pressures[45, 46] or at low concentrations and low pressures.[23] In both cases, the volume flow rate is proportional to the applied pressure. The factor $K$ in its original form describes the geometrical form of the pore. However, given that the fluid contains vesicles that need to be pushed though the pore, Bruinsma[45] converted it into an effective permeability which is given by:

$$K_{eff} = \frac{N \pi R_p^4}{8 + 0.233(nL^*)\left( \frac{L^*}{R} \right)} \tag{2}$$

Where $N$ is the number of pores in the membrane, $R_p$ is the average radius of the pores, $n$ is the number of vesicles in a pore of length $L^*$ and $R$ is an effective radius of the pore that is dependent on the velocity of the fluid moving through this pore. As the fluid moves though the pore, a lubrication layer forms along the side of the pore which increases in thickness as the velocity of the fluid increases.[45] This phenomenon is given by $R = R_p - h^*(v)$ where $h^*$ is the thickness of the lubrication layer and $v$ the velocity of the fluid.

Based on the polydispersity of the extruded vesicles, the mechanism by which the vesicles are formed is pressure- rather than flow rate-dependent.[23, 47] However, flow rate has also been observed to be linked to the size of the extruded vesicles.[46] Studies of emulsions and lamellar phases exposed to shear by Diat[48] and Mason and Bibette[49] showed increased shear leading to smaller droplets. Hunter et al.[46] explain this effect via the layer of lubrication that forms on the inside of the pore as the solution is extruded. The higher the flow rate, the thicker this lubrication layer, the smaller the effective pore size. It is well documented, that the main factors determining the size of the extruded vesicles is the pore size.[16, 50-52]

The diameter of the resulting vesicle, especially during the first extrusion pass, is determined when the large vesicle enters the pores of the membrane.[23] As more and more vesicles reach necessary size to easily pass through the pores of the membrane, the pressure needed to press the vesicles through the membrane continually decreases.[23] The lower the extrusion pressure, the larger the resulting vesicle.[23] In reverse, studies using extruders featuring constant extrusion pressures find an increase in flow rate which eventually plateaus once all vesicles are small enough to easily pass through the membrane pores.[46]

The pressure needed to for a vesicle to move through the pore is equal to the lysis tension of this vesicle because the passage through the pore causes the vesicle to rupture.[46] The Laplace relation gives the pressure difference necessary to break the surface tension of an interface, depending on the curvature of that interface.[46, 53]

$$\Delta P = 2\gamma H \tag{3}$$

All variables apply to the interface: $\Delta P$ is the pressure difference across the interface, $\gamma$ is the surface tension and  is the mean curvature. Based on this relation, Hunter et al. expanded the equation to take into account the radii of the vesicle being extruded and of the membrane pore[46]

$$P_1 - P_0 = 2\gamma \left[ \frac{1}{R_p} - \frac{1}{R_o} \right] \approx 2\gamma \left( \frac{1}{R_p} \right) \tag{4}$$

Here, $P_1$ is the pressure applied to the solution, $P_0$ is the atmospheric pressure, $R_p$ is the radius of a vesicle small enough to pass through the pore and $R_o$ is the radius of the original larger vesicle being pushed against the pore entrance.

## MATERIALS AND METHODS

### Preparation of Liposomes

1,2-Dioleoyl-sn-glycero-3-phosphocholine (DOPC) was purchased from NOF American Corporation. 54 mM DOPC stock solutions were prepared in chloroform. 0.5 mL aliquots of DOPC stock solutions were then evaporated under a $N_2$-stream and the resulting lipid cakes stored under vacuum overnight. The dried lipid cakes were then rehydrated with 2 mL of ultrapure water (18.2 MΩ/cm) and vortexed. The samples were then exposed to four freeze-thaw cycles. Each cycle consisted of 10 min at −20 °C in a freezer and 10 min at 50 °C on a hot plate. The samples were then extruded at room temperature. For the extrusion process, 100 nm polycarbonate membranes were used.

### Dynamic Light Scattering (DLS)

Dynamic Light Scattering (DLS) was performed using a Malvern Zetasizer Nano ZS equipped with a 633 nm, 30 mW He-Ne laser at a scattering angle of θ = 173° at a temperature of 25 °C and an equilibration time of 60 s. Hydrodynamic diameter, ($D_H$) and polydispersity ($PDI$) were determined. First, the dynamic correlation function was measured, from which the Malvern Zetasizer Nano ZS instrumental software was used to determine the $D_H$ and $PDI$ based on cumulants' analysis using an exponential decay function.[54] Each data point represents three averaged samples. Each sample has been measured three times and the values combined to form an average. Error bars shown in the figures represent the standard deviation.

## AUTOMATIC EXTRUDER SETUP

The automatic extruder was designed for student researchers and features a simple setup. The project is assumed to fit within the limited numbers of hours a typical student researcher can spend in a research laboratory. Once completed, the project can be further expanded upon. Hence, the project can be adapted to the background of the students. Readers are encouraged to upload software and hardware improvements, including technical drawings to our GitHub archive intended to make the project freely available. The link can be found HERE.

Two syringe pumps are synchronized using the LabVIEW software. Implementing the software will afford the students insight into the backend setup of scientific instruments while the use of the completed automatic extruder will provide them with knowledge about macroscopic processes, self-assembly, and fluid dynamics. Building a base for the syringe pumps or adding additional heating functions to the synchronized pumps will allow student to acquire skills related to the design and application of mechanical parts.

*Necessary Components*

The following parts and components must be available in order to set up the automatic extruder described in this project:

- 1 or 2 syringe pumps from New Era Syringe Pump Inc.:
- Model number 1000, 12 Volts direct current, 0.75 amperes.
- 1 Avanti Mini-Extruder
- 2 1 mL gastight glass syringes with removable needles and an internal diameter of 4.61 mm. (Part of the mini-extruder kit available from Avanti Polar Lipids)
- 2×10 mm Filter supports (per 1 mL extruded solution)
- 1×0.1 μm Etched Polycarbonate Membrane (per 1 mL extruded solution)
- 1 computer
- 2 RS-232 Pump Network (CBL-PC-PUMP) connector cables
- 2 Pump Power cables
- 2 network cables
- Graphical user interface[30]

*Syringe pump setup*

Two automatic single syringe pumps (New Era Pump Systems, Inc.) were fastened to a wooden board, syringe opening sides facing each other. A distance of 12 cm between the individual syringe pumps allowed for one Avanti Mini-extruder, loaded with two 1 mL Hamilton® gastight glass syringes to be placed in the syringe pump system (**Figure 1**). The two NE-50X syringe pumps are set up to communicate with one another in order synchronize the pumping process. One pump is programmed as the "master" pump. Any configurable variable applied to the master pump will also be applied to the "secondary" syringe pump. Each pump is programmed with an address (typically 0 and 1, 0 being the master pump and 1 being the follower pump) in order to be differentiated by the computer, which will only communicate with the master pump. The two pumps communicate using a regular CAT-5 ethernet network cable connected to the back of the syringe pumps. An RS-232 to USB adapter connects the back of each pump (RS-232) to the computer (USB) running the LabVIEW program.

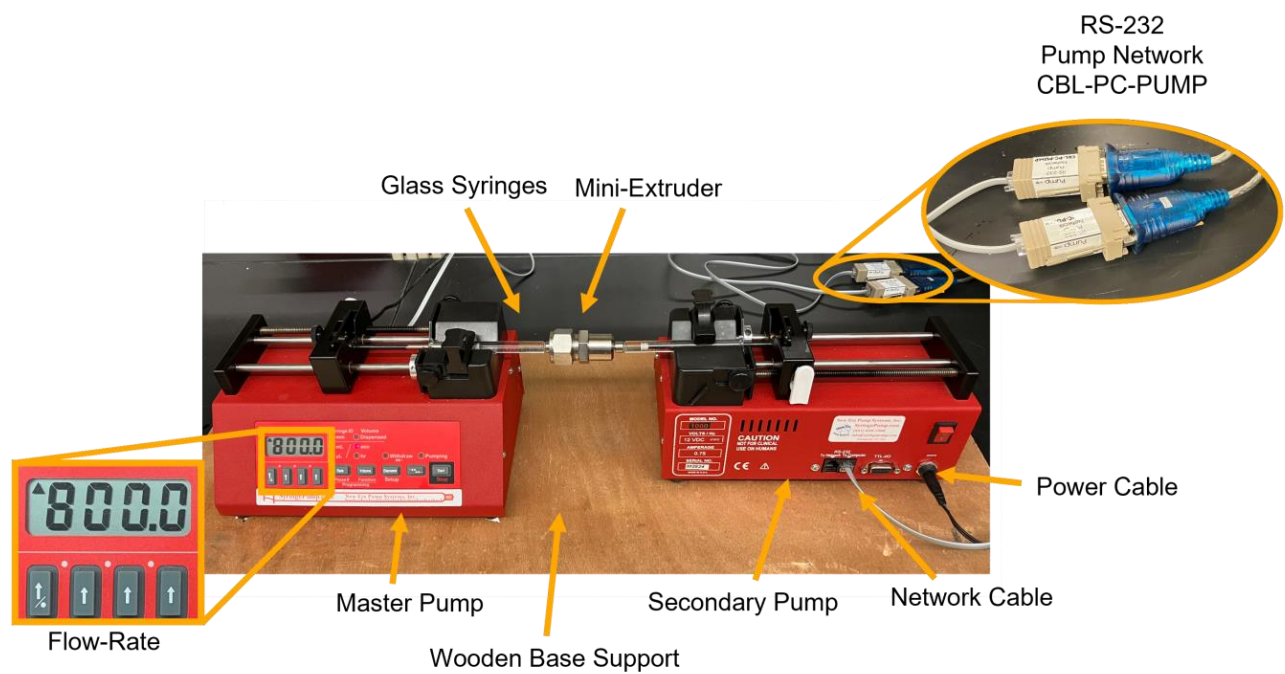

**Figure 1:** Labeled image of the automatic extruder using two connected automatic syringe pumps.

*Software Development*

Engineering projects necessitate careful decisions throughout the project. Workflow diagrams can help assure that the efficiency of the project is systematically and thoroughly challenged with potential improvements and optimizations. A diagram such as the one shown in **Figure 2** can be used to lead students on the decision-making process.

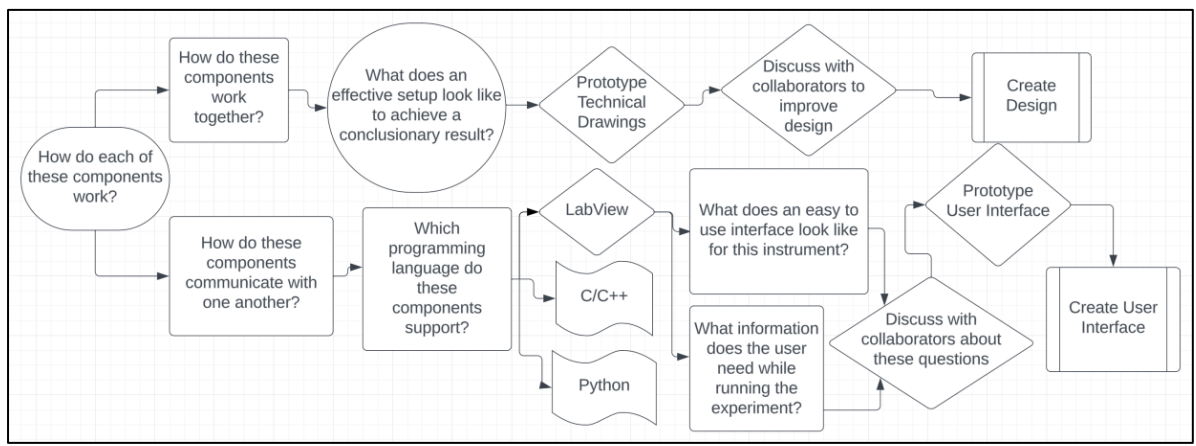

**Figure 2:** Workflow diagram showing questions and decisions made throughout the project to achieve an optimal and effective instrument.

*LabVIEW*

The LabVIEW workflow file available in the supplementary information shows LabVIEW's block chart for this project. In all LabVIEW applications there is a starting set of variables collected from a resource. In this case, the initial resource is a driver software downloaded from the National Instruments page for the syringe pumps. Once these variables are ready to be used, they can be imported into the SubVI's. SubVIs are repeatable LabVIEW programs within a larger LabVIEW program. They allow the software to become more organized and reduce software development time. Referring to the workflow file provided in the supplementary information: Section A shows a series of SubVIs whose objectives are to set presets to the two syringe pumps before operation has started. Sction B shows the repeated piece of code during the run time of the software. This is LabVIEW's equivalent translation to a "For Loop" and behaves similarly to other software languages. Information from within this loop status is constantly sent to the front panel to assure the user that the instrument is operating correctly throughout the duration of the run time. Section C is the following loop primarily checking for the additional commands from the user during the run time of the instrument. This piece of the code also handles any error codes that may arise throughout the runtime and shuts down the instrument in the case of an error code. On the left beginning side of the block chart, the initialization of the syringe pumps and configuration settings are sent to the pump. Following this chain of commands, a "For" command block is presented to run the given configuration until it is complete, or the stop command is pressed on the front panel of the LabVIEW software.

The SubVI's within the LabVIEW block chart are designed in a way to be interchangeable with different pumps and their respective drivers. The drivers presented with the software will work with a large set of pump instruments.

Graphical User Interface (GUI)

A graphical user interface (GUI) can be created using LabVIEW (**Figure 3**). The essential part of the GUI is the Pump Control panel that allows to select important parameters, such as flow rate, solution volume and number of extrusions to perform as well as the inner diameter of the syringes used and the initial direction of extrusion. The hardware limits on the syringe pump are stored in a separate configuration file. The flow rate set through the software is displayed in the display on the user interface of the master pump (**Figure 1**).

This version of the software assumes the two-syringe pump setup illustrated by the Configuration Settings panel (left) that shows a master and a secondary pump. To connect it with the computer a USB to serial adapter as installed and the ports COM3 and COM4 assigned to the respective USB ports. Hence, the PC with Windows 11 operating system and the syringe pumps are connected by a serial (RS-232) interface that communicates with a baud rate of 19200 baud.

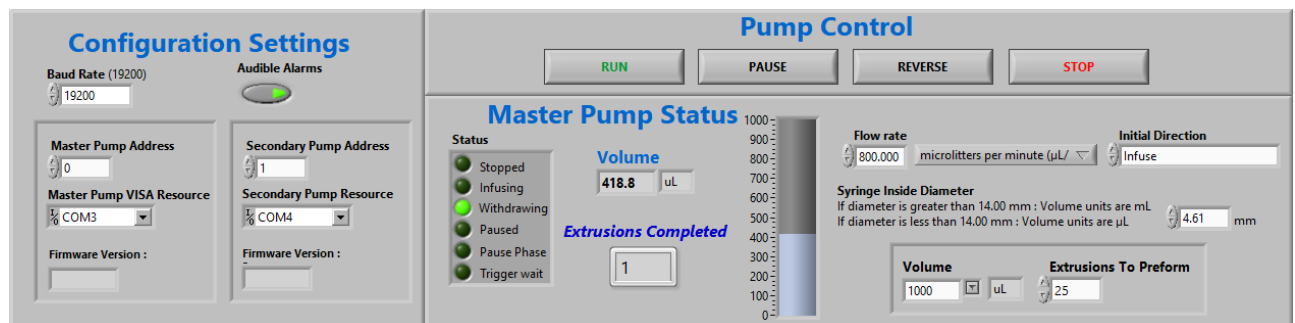

**Figure 3:** Screenshot of the LabVIEW control panel for the two-pump automatic extruder software. The configuration settings on the right show the connection to the computer while the pump control is used to set parameters for the extrusion that will be executed. The link to the GitHub can be found HERE.

**TESTING THE AUTOMATIC EXTRUDER**

The automatic extrusion system was compared to manual extrusion according to the number of passes, the flow rate, and a stepwise change in flow rate. The differences in hydrodynamic diameter $D_H$ and *PDI* between manual extrusion by multiple experimenters and the automatic extrusion were evaluated. Unless stated otherwise, each datapoint consists of three separately prepared samples whose $D_H$ was analyzed via three averaged runs of dynamic light scattering.

*Influence of Number of Passes*

The number of passes through the membrane influences the size of the resulting vesicles.[55] The automatic extruder uses this system's maximum flow rate of 800 µL/min. Manual extrusion by the authors was found to average about 1000 µL/min. The extrusions were performed using freshly prepared samples at ambient temperature. About 5% of the lipid mass are lost via extrusion decreasing slightly with the number of passes (**Figure SI 2**). With the exception of **Figure 5**, all values marked as "manual extrusion" include samples prepared by students 2 and 3. Data by student 1 was taken from an earlier work[55] by our group and not included in the averages as different numbers of passes were used. All samples were extruded for an odd number of passes to prevent any larger impurities in the initial sample from remaining after the extrusion.

Automatically and manually extruded samples decrease in $D_H$ with increasing number of passes until the lowest $D_H$ is reached (**Figure 4a**). Exponential decay functions are added as a guide to the eye. The automatic extruder seems to produce liposomes with slightly larger diameter. The *PDI* of both automatic and manually extruded samples first decreases and then seem to remain constant. The *PDI* of the manual extrusion remains below that of the automatic extrusion and eventually become the same size after 40 extrusions (**Figure 4b**).

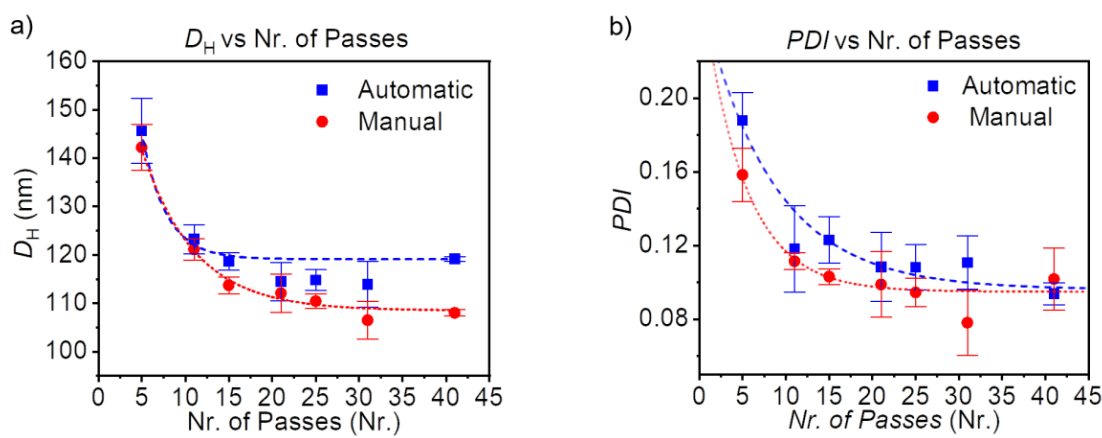

**Figure 4:** Influence of the number of passes for the automatic and manual extrusion on the a) $D_H$ and b) *PDI* of DOPC liposomes. Lines are guides to the eye. All samples were extruded at ambient temperature (23 °C). The automatic extruder samples were extruded at a flow rate of 800 µL/min.

Since larger vesicles have a smaller surface tension (**equation 2**), the lower the vesicle radius, the larger the lysis tension needed to rupture the vesicle.[45] Considering the automatic extruder working based on constant flow rate rather than constant pressure, as fewer large vesicles remain in solution, the pressure built up by the plunger decreases. Because this pressure no longer reaches the lysis tension of the smaller vesicles, the vesicles remain slightly larger than those prepared by manual extrusion. During manual extrusion, the experimenter naturally operates at constant pressure rather than constant flow rate, causing the applied pressure to remain above the lysis tension for the vesicles for a longer time leading to smaller vesicles.

This effect may be the reason for the stagnation in the vesicle diameter decrease during automatic extrusion (**Figure 4a**). It should also be considered though, that the main reason for the constant diameter in the best fit line may be the final value at 41 extrusion passes. Considering the error bars of both the 35 passes and 41 passes diameter values. The 41 passes diameter value falls just outside the 35 passes diameter value while previous values fell continuously within each other's error bars. This may suggest that the 41 passes value should be considered an outlier due to an impurity and excluded from analysis. Excluding the 41 passes diameter value would cause the diameter value to remain constant at a slightly higher number of passes. It would also mirror the manual extrusion more closely.

A similar trend can be observed in the *PDI* values of the of the vesicles while increasing the number of extrusion passes (**Figure 4b**). The *PDI* decreases with increasing number of passes and this decrease is slightly faster for the manual extrusion than the automatic extrusion. At the maximum number of extrusion passes, both extrusion methods reach a similar *PDI*. Again, the unusually low error of the 41 passes sample should be considered. If this data point were removed, as discussed for the $D_H$, the *PDI* curves would match the $D_H$ curves more closely. $D_H$ and *PDI* values for the automatic extrusion remain slightly above those achieved by manual extrusion. Given the reasoning above, manual extrusion may be able to increase pressure beyond that of the automatic extruder, leading to a lower minimum diameter for the vesicle than is achievable by keeping the flow rate constant and thereby causing a decrease in the pressure as is the case for automatic extrusion. Given the reasoning above, manual extrusion may be able to increase pressure beyond that of the automatic extruder, leading to a lower minimum diameter for the vesicle than is achievable by keeping the flow rate constant and thereby causing a decrease in the pressure as is the case for automatic extrusion.

In order to understand the effect of different student's manual extrusion on the vesicle diameter, data from different students was compared to the automatically extruded samples (**Figure 5**). The figures show the $D_H$ and *PDI* of extruded samples as a function of the number of extrusions by different students and the automatic extruder. For student 1, all standard deviations fell within the space of the symbols. Error bars were therefore omitted. For all students, the $D_H$ fell below the value achieved using the automatic extruder (**Figure 5a**). With the exception of the 41 passes value, this is also the case for the *PDI* (**Figure 5b**). This result corroborates the earlier suggestion that the natural increase of pressure by several different students resulted in a lower minimum diameter than could be achieved by the automatic extrusion.

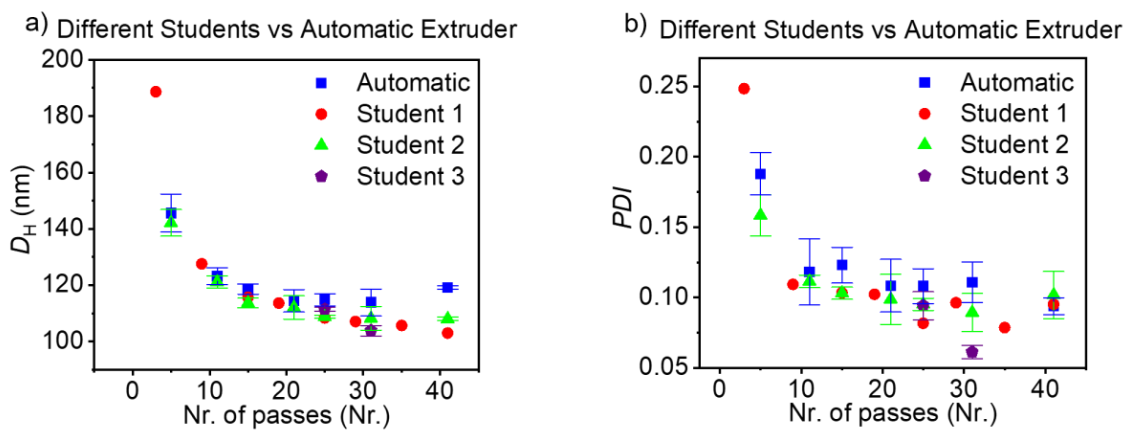

**Figure 5:** Comparison of data extruded by different students. For data by student 1, no standard deviation data was available. The data was digitized from an earlier work.[55] a) shows $D_H$ as a function of the number of passes and b) the *PDI* as a function of the number of passes. The flow rate of the automatic extruder was 800 µL/min. All extrusions were performed at ambient temperature (23 °C).

**Figure 6** shows the standard deviation of the $D_H$ and *PDI* for each sample as a function of the number of passes. The data shows an exponential decrease in standard deviation for the $D_H$ values and a linear decrease for the *PDI* values. While $D_H$ values decrease along a similar trend, *PDI* values decrease to a greater degree in automatic than in manual extrusion. Red dotted lines mark the manual and blue dashed the automatic values. No error bars are shown for this figure since the single value includes all data samples.

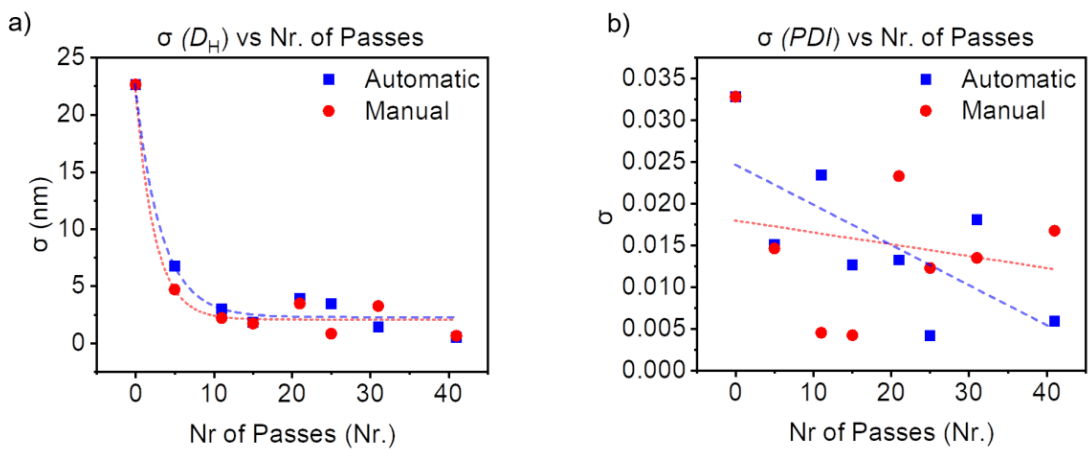

**Figure 6:** Comparison of the standard deviation of a) $D_H$ and b) *PDI*. Highlighted in grey are the samples that were prepared 10 times in order to test reproducibility. In b) 21 passes using the automatic extruder was calculated using 9 samples, as one sample was contaminated resulting in a much larger value. All samples were extruded at ambient temperature (23 °C). The automatic extruder samples were extruded at a flow rate of 800 µL/min.

In terms of reproducibility, the automatic extruder seems to be slightly better in terms of $D_H$. For most $D_H$ and *PDI* values, the standard deviation between samples extruded for an equal number of passes is slightly smaller for the automatic extrusion than the manual one (**Figure 6**). Nonetheless, while the automatic extruder is limited in terms of the minimum vesicle diameter, the constant flow rate and consequently the constant change in the pressure throughout the extrusion process leads to slightly better reproducibility using the automatic extruder.

*Influence of Flow Rates*

The flow rates are investigated using only the automatic extruder, as constant flow rates are difficult to achieve by a human experimenter. Each flow rate sample was extruded for a total of 21 passes as, in the previous section, this was found the be the number of passes at which the vesicle $D_H$ remains constant. **Figure 7** shows $D_H$ and *PDI* of the extruded liposomes as a function of the flow rate used for their extrusion. There appears to be a weak dependence of the $D_H$ on the flow rate (**Figure 7a**) and a slightly stronger one for the *PDI* (**Figure 7b**). While flow rates show a slight correlation with liposome diameter. The more visible trend is observed for the *PDI*. Here, the highest flow rate of 800 µL/min shows a *PDI* of around 0.09 while a flow rate of 400 µL/min resulted in around 0.14.

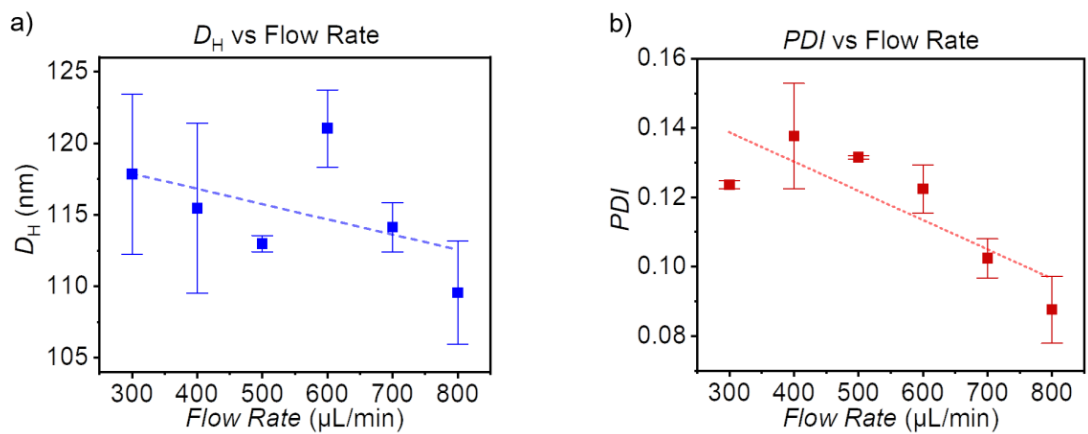

**Figure 7:** Influence of the flow rate on a) $D_H$ and b) *PDI* of the DOPC liposomes extruded automatically with varying flow rates. The horizontal blue line marks the average liposome diameter.

Increasing the flow rate seemed to have some effect on the minimum achievable vesicle diameter using the automatic extruder. This contrasts with findings from the literature, where a doubled flow rate did not affect the size of the vesicles.[23] However, given the increase in the lubrication layer with increased flow rate described by Bruinsma,[45] a slight decrease in diameter would be expected. A small dependence of the $D_H$ was found in the data. The *PDI*, on the other hand, clearly decreased with increasing flow rate (**Figure 7**), counter to Frisken et al. who found that the *PDI* was not connected to any of the variables of the extrusion process. They suggested that this may be due to the fact that extrusion could be considered a type of fragmentation process, for which these large size distributions would be expected.[23] If the effect visible in the data presented here is indeed due to the decreased diameter caused by the lubrication layer, smaller fragments and a lower *PDI* would be expected.

*Influence of a Stepwise Flow Rate Adjustment*

During the testing of stepwise flow rate changes, the flow rate was increased by a different number of steps over a total of 29 passes. The flow rates and number of passes can be seen in **Figure 8**. Stepwise changes of the flow rates are implemented in order to simulate the increased flow rates during manual extrusion. The resistance of the solution decreases as it passes through the membrane and is automatically adjusted for by a human experimenter.

Three sets of stepwise flow profiles were selected. For the first sample, 15 extrusion passes were performed at a flow rate of 500 µL/min followed by 14 extrusion passes at 800 µL/min. As two flow rates resulting in two initial extrusion pressures were involved in this sample, it was named P2. Sample P3 included three different flow rates and sample P4 included four different flow rates. **Figure 8** graphically displays the change in flow rate. Finally, an additional flow rate sample (P29) was added, for which the flow rate was increased by 11 µL/min with every pass.

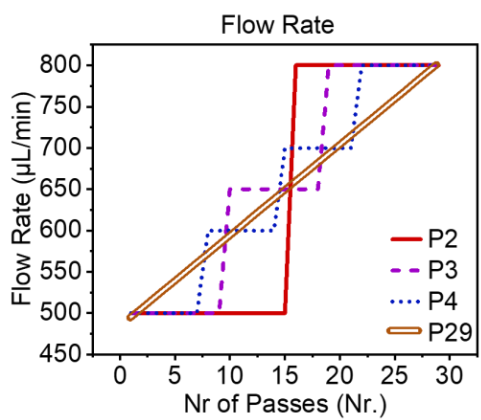

**Figure 8:** The number of passes at each flow rate for samples P2-4 displayed graphically.

As the manual extrusion has so far been shown to result in smaller, more uniform particles, the automatic extrusion procedure was adjusted so as to mimic the increase in flow rate that results from the decrease in particle size during the manual extrusion process.[23, 47]    The results of the stepwise increase in flow rate are shown incorporated into the figures depicting the effect of the number of passes in order to allow for evaluation of the stepwise increase in flow rate (**Figure 9**). However, changing the flow rates in a stepwise manner had no positive effect on the diameters of *PDI* of the liposomes. In fact, both *PDI* and diameter placed between or above both trendlines, with the exception of the *PDI* of sample P29.

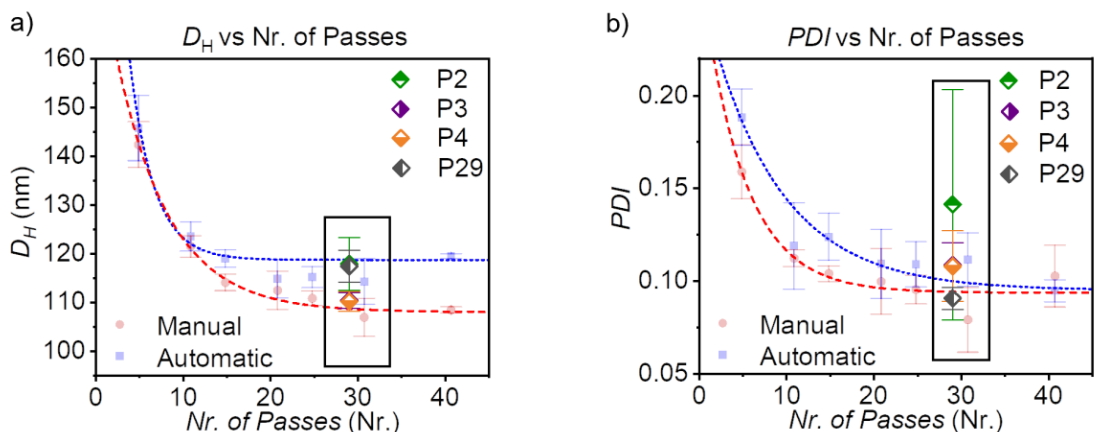

**Figure 9:** Results of the stepwise flow rate trials incorporated into the greyed-out figures depicting the effect of the number of extrusion passes on the $D_H$ (a) and *PDI* (b) of the DOPC liposomes. P2 is shown in green, P3 in purple and P4 in orange.

The stepwise increase to mimic the increasing pressure during manual extrusion also showed no effect (**Figure 9**). Instead, all $D_H$ values for the 29 passes used for this experiment fell within the error bars of the 35 extrusion passes automatic extruder sample. As the 35 passes sample was one of the datapoints for which reproducibility was tested by averaging 10 samples, the samples for which stepwise increase in the flow rate was implemented seem to only have flushed out the variability of the $D_H$ for the automatic extruder rather than brought it closer to the minimum diameter achievable by manual extrusion.

**PERSPECTIVES**

This work presents a simple method for students to come into contact with an important self-assembly process for that can be used for an introduction to macromolecular chemistry. Preparation of liposomes can be a valuable tool to connect in-lab experiments to real-life applications. Manual extrusion in particular gives students the opportunity to learn about the importance of the variation of only a single variable in terms of the scientific method. Allowing different students to extrude liposome samples results in different vesicle diameters due to the change of a number of variables associated with individual experimenters. The use of an automatic extruder limits the number of changing variables while also excluding ones that may have been favorable to achieving low diameter vesicles. The flow rate-based extrusion system cannot be adjusted to mimic manual extrusion by adjusting the flow rate throughout the process. The extruder is, however, very useful when applied to projects necessitating large throughput where a slightly larger vesicle diameter is acceptable. In addition, the time gained by using a simple automatic extruder can be used to convey the theoretical knowledge needed. This way, class time can be used more efficiently.

The project described here supplies instructions on how to build an automatic extruder and offers suggestions as to complementary skills that may be helpful in an industrial laboratory. This portion specifically can be adapted to the background of the students, be it in mechanical engineering, software engineering, chemistry, or other scientific or engineering field, using the type or automatic extruder. Through the LabVIEW Program, all that needs to be present in an undergraduate lab to build an automatic extruder are two syringe pumps and a computer. It is likely that the university can supply the instrument for size analysis. This also offers an opportunity for interdisciplinary cooperation between students of different departments.

To further simplify the setup and make it more suitable for an undergraduate laboratory environment, a Raspberry PI or tablet could be implemented to make operation more intuitive. Additionally, the syringe pumps may be adapted to be operated in a pressure sensitive mode, which may lead to lower vesicle diameters and would allow further discussion of the theory involved in vesicle preparation by extrusion.

**ASSOCIATED CONTENT**

Supporting Information

The Supporting Information is available on the ACS Publications website at DOI:

LabVIEW workflow file (PDF)

Supporting Information (PDF)

Github Link https://github.com/gschneidergroup/Simplistic-Automatic-Extruder

**AUTHOR INFORMATION**

Corresponding Author

Gerald J. Schneider – Department of Chemistry and Department of Physics, Louisiana State University – Baton Rouge, Louisiana 70803, United States; 0000-0002-5577-9328; Email: gjschneider@lsu.edu

**ACKNOWLEDGMENTS**

B J, and G J S gratefully acknowledge funding by the U.S. Department of Energy (DoE) under Grant DESC0019050. Further, G J S, D G, E B, A D and S G gratefully acknowledges funding by the U.S. National Science Foundation (NSF) under Award Number 1808059. Finally, S K and G J S gratefully acknowledge funding by the American Chemical Society Petroleum Research Fund (ACS PRF) under the grant number 61982-ND7. In addition, we would like to thank our collaborators at the NIST Center for Neutron Research (NCNR) for their help and advice.

**SUPPLEMENTARY INFORMATION**

*SI 1: Influence of Water on Liposomes: Deionized vs Ultrapure*

The influence of the type of water of water on the liposomes' hydrodynamic diameter ($D_H$) and polydispersity ($PDI$) was tested. Three additional samples of liposomes were prepared using deionized water from the wall faucet in the laboratory. These were compared with the results on liposomes prepared in ultrapure water (18.2 MΩ/cm) of the main text. For the liposomes in deionized water the same preparation using the automatic extruder (flow rate of 800 μL/min through a 100 nm polycarbonate membrane extruded for 21 passes) was chosen. **Figure S1** shows that there is no change in the diameter due to the change in water source within the accuracy of the experiment.

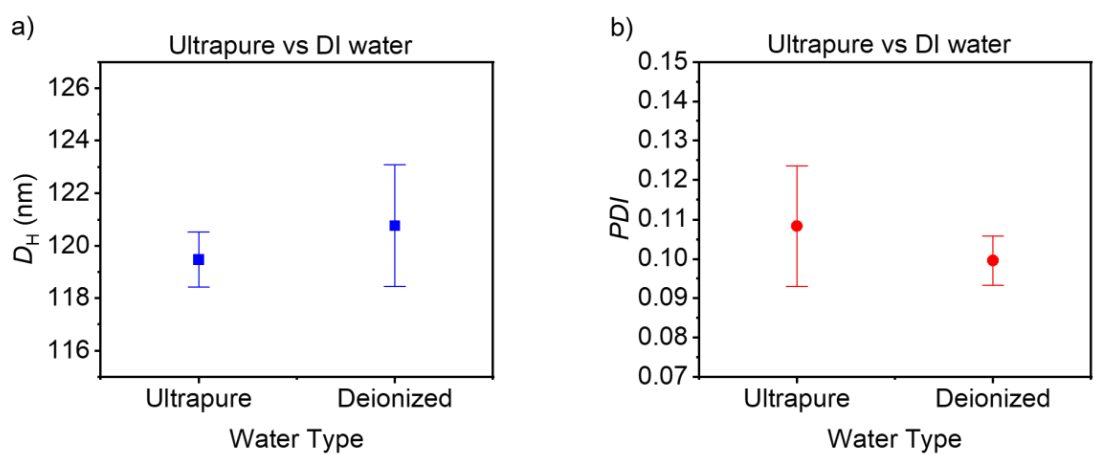

**Figure S 1:** Liposome a) hydrodynamic diameter ($D_H$) and b) polydispersity ($PDI$) for samples prepared using deionized water from the laboratory faucet and using a Barnstead Smart2Pure filtration system operating at a resistivity of 18.2MΩ/cm.

*SI 2: Sample loss during extrusion*

The mass or the polycarbonate membrane and the two filter supports was recorded before extrusion of the samples. Three different numbers of extrusion passes 5, 11, and 21 passes were tested (**Figure S2**). For each number of passes, three samples were prepared. Each was extruded using the automatic extruder at 800 μL/min through a 100 nm polycarbonate membrane. The used polycarbonate membranes and filter supports were then dried overnight, and their mass recorded the next day. All samples showed an average loss of about 6% of sample mass remaining in the polycarbonate membrane and filter supports after extrusion. While the average value decreases, there is no statistical difference. The values are within each other's error bars.

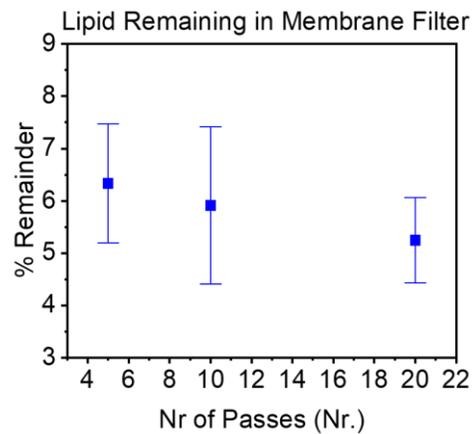

**Figure S 2:** Lipid remaining in the polycarbonate membrane after automatic extrusion at 800 μL/min.